\documentclass[%
aip,
amsmath,amssymb,
reprint,%
]{revtex4-1}

\usepackage{graphicx}               
\usepackage{dcolumn}                
\usepackage{bm}                     

\usepackage[utf8]{inputenc}
\usepackage[T1]{fontenc}
\usepackage{amsmath}
\usepackage{amssymb}
\usepackage{mathptmx}
\usepackage{braket}
\usepackage{longtable}
\usepackage{siunitx}
\usepackage{etoolbox}
\usepackage{gensymb}
\usepackage{comment}
\usepackage{natbib}

\makeatletter
\def\@email#1#2{%
 \endgroup
 \patchcmd{\titleblock@produce}
  {\frontmatter@RRAPformat}
  {\frontmatter@RRAPformat{\produce@RRAP{*#1\href{mailto:#2}{#2}}}\frontmatter@RRAPformat}
  {}{}
}%
\makeatother

\DeclareSIUnit[number-unit-product = {\,}]\cal{cal}
\DeclareSIUnit[number-unit-product = {\,}]\mol{mol}

\usepackage{xcolor}
\usepackage[normalem]{ulem}

\newcommand{\trial}{\mathrm{T}}
\newcommand{\init}{\mathrm{I}}
\newcommand{\guide}{\mathrm{T}}
\newcommand{\loc}{\mathrm{L}}
\newcommand{\noci}{{N_\mathrm{d}}}
\newcommand{\numdet}{\noci}
\newcommand{\epochs}{N_\xi}
\newcommand{\epoch}{\xi}

\begin{document}

\preprint{AIP/123-QED}

\title[]{Self-Refinement of Auxiliary-Field Quantum Monte Carlo via Non-Orthogonal Configuration Interaction}

\author{Zoran Sukurma}
\affiliation{University of Vienna, Faculty of Physics, Kolingasse 14-16, A-1090 Vienna, Austria}
\affiliation{University of Vienna, Faculty of Physics \& Vienna Doctoral School in Physics,  Boltzmanngasse 5, A-1090 Vienna, Austria}
\author{Martin Schlipf}
\affiliation{VASP Software GmbH, Sensengasse 8, 1090 Vienna, Austria}
\author{Georg Kresse}
\affiliation{University of Vienna, Faculty of Physics and Center for Computational Materials Science, Kolingasse 14-16, A-1090 Vienna, Austria}
\affiliation{VASP Software GmbH, Sensengasse 8, 1090 Vienna, Austria}

\date{\today}

\begin{abstract}
For optimal accuracy, auxiliary-field quantum Monte Carlo (AFQMC) requires trial states consisting of multiple Slater determinants.
We develop an efficient algorithm to select the determinants from an AFQMC random walk eliminating the need for other methods.
When determinants contribute significantly to the non-orthogonal configuration interaction energy, we include them in the trial state.
These refined trial wave functions significantly reduce the phaseless bias and sampling variance of the local energy estimator.
With 100 to 200 determinants, we lower the error of AFQMC by up to a factor of 10 for second row elements that are not accurately described with a Hartree-Fock trial wave function.
For the HEAT set, we improve the average error to within the chemical accuracy.
For benzene, the largest studied system, we reduce AFQMC error by 80\%  with 214 Slater determinants and find a 10-fold increase of the time to solution.
We show that the remaining error of the method prevails in systems with static correlation or strong spin contamination.
\end{abstract}

\maketitle

\section{Introduction}
\label{sec:intro}

Accurate large-scale correlation-consistent calculations have become feasible with advances in computing power.\cite{BenaliDMCL72020,AlHamdaniDMC-CCSDT2021,SchaeferCCSDcT2024}
Coupled-cluster singles, doubles, and perturbative triples (CCSD(T))\cite{Cizek1966,CizekPaldus1971} and fixed-node diffusion Monte Carlo (FN-DMC)\cite{Anderson1976DMC,Ceperley1977DMC,Foulkes2001QMC} are two of the most prominent methods, offering an excellent balance between accuracy and computational cost.
However, recent studies\cite{BenaliDMCL72020,AlHamdaniDMC-CCSDT2021} indicate increasing discrepancies between the two methods for large non-covalent complexes. 
For example, the discrepancy for a C$_{60}$ buckyball inside a [6]-cycloparaphenyleneacetylene ring (132 atoms) amounts to \qty{7.6}{\kilo \cal / \mol} after ruling out all stochastic uncertainties.
Such large discrepancies lead to mistrust in both methods, especially as each is considered highly reliable for predicting non-covalent interaction energies.
Sch\"afer \emph{et al.}\cite{SchaeferCCSDcT2024} pointed out that CCSD(T) may not be sufficiently accurate and that correcting the perturbative triples\cite{MasiosCCSDcT2023} improves interaction energies at the cost of less accurate total energies.

Auxiliary-field quantum Monte Carlo (AFQMC) \cite{Zhang2003Phaseless,Motta2018AFQMCREVIEW,Motta2019Review} has emerged as a promising alternative to CCSD(T), gaining increasing popularity in recent years.\cite{Purwanto2011HCa,Virgus2012CobaltGraphene,Purwanto2105ChromiumDimer,Motta2017Hchain, Motta2109-AFQMC-ED, Shee3dTrans2019,Lee2020BenzeneAFQMC,Lee2020SymmetryBreaking,Rudhsteyn2022Metallocene,Lee2022_20yearsAFQMC,sukurmaHEAT2023,Amir2023,WeiAFQMCBench2024,MahajanCISD-AFQMC2024}
Several key factors contribute to the growing popularity of AFQMC:
(i) its formulation in second quantization, enabling straightforward application on top of the mean-field methods,
(ii) its favorable computational scaling---$\mathcal{O}(N^4)$ for AFQMC compared to $\mathcal{O}(N^7)$ for CCSD(T), and
(iii) its demonstrated accuracy.
In its simplest formulation with a single-determinant trial wave function, AFQMC accuracy typically falls between that of CCSD and CCSD(T).\cite{Shee3dTrans2019,Lee2022_20yearsAFQMC,sukurmaHEAT2023,WeiAFQMCBench2024}
The main source of error in AFQMC is the phaseless approximation\cite{Zhang2003Phaseless,Motta2018AFQMCREVIEW,sukurmaHEAT2023,WeberAFQMCconstr2023,XiaoAFQMCconstr2023} introduced to control the fermionic phase problem.
Additionally, AFQMC requires many steps to reduce the sampling variance leading to a large prefactor of the quartic scaling.
Together, these factors continue to hinder efficient large-scale AFQMC calculations.

More accurate trial wave functions reduce the phaseless error and sampling variance and are thus a promising route towards more accurate AFQMC.
Particle-hole multi-Slater determinants (PHMSD) are the most convenient choice for constructing such trial wave functions.
Since the computational cost of a naive implementation scales linearly with the number of determinants, many groups have developed fast local energy evaluation methods for PHMSD trial wave functions.\cite{Shee2018AFQMC-GPU,MahajanFastLocen2020,Shi2021RecentDevAFQMC,Mahajan2021TamingSignProblem,Mahajan2022AFQMCsCI,MahajanCISD-AFQMC2024}
Shee \emph{et al.}\cite{Shee2018AFQMC-GPU} were the first to report an efficient evaluation of local energies over PHMSD trial wave functions using the Sherman-Morrison-Woodbury formula\cite{GolubLoanMC}.
Later, Shi and Zhang\cite{Shi2021RecentDevAFQMC} optimized the algorithm further, performing calculations with 10,000 determinants and achieving a 60-fold speedup compared to the naive implementation.
Their algorithm scales as $\mathcal{O}(N_gN^2N_e\tilde{N}_d)$, where
$N_g$ is the number of auxiliary fields,
$N$ is the number of orbitals,
$N_e$ is the number of electrons, and 
$N_d$ is the number of determinants.
The notation $\tilde{N}_d$ indicates sublinear scaling with the number of determinants.
Mahajan \emph{et al.}\cite{MahajanFastLocen2020,Mahajan2021TamingSignProblem} applied the generalized Wick's theorem\cite{Shavitt_Bartlett_2009} to design an algorithm that scales as $\mathcal{O}(N_gN^2N_e + N_gN_d)$.
They reported calculations using 10,000 determinants with only a threefold increase in computational cost compared to the single-determinant case.\cite{Mahajan2022AFQMCsCI}
Configuration-interaction-singles-doubles (CISD) trial wave functions containing up to {half a billion of PHMSDs represent the largest determinantal expansion used with AFQMC to date.\cite{MahajanCISD-AFQMC2024}

In addition to the PHMSD wave functions, non-orthogonal multi-Slater determinants (NOMSD) provide an appealing alternative, often requiring fewer determinants than the orthogonal case.
Borda \emph{et al.}\cite{Borda2019NONSD} performed AFQMC calculations using up to 250 NOMSD determinants.
They generated the trial wave functions using the few-determinant approach \cite{SchmidFEDOrig1988,SchmidFEDRev2004,JimenezHoyosPHF2013} and the resonating Hartree-Fock method.\cite{FukutomeResHF1988,IkawaResHFOpt1993}
For these wave functions, the local energy evaluation scales linearly with the number of determinants, which limits its practical application as the number of determinants grows.
Alternative choices for the trial state of AFQMC simulations are general Hartree-Fock (GHF) wave functions,\cite{DanilovGHFAFQMC2024} matrix-product states,\cite{JiangMPSAFQMC2024} and states obtained using quantum computers.\cite{Huggins2022qcAFQMC,AmslerQC-CL-AFQMC2023,KiserQCAFQMC2024,KiserCSQC-AFQMC2024}

In this work, we use the AFQMC method itself to generate candidate determinants for the non-orthogonal Configuration interaction (NOCI) and refine the AFQMC trial wave functions.
We carefully design the selection process to achieve the most compact NOCI expansion.
During each epoch of determinant selection, we perform a short AFQMC random walk and select determinants that
(1) have sufficiently negative local energies,
(2) exhibit a small overlap with the currently selected NOCI wave function, and
(3) contribute significantly to the correlation energy.
We demonstrate that the resulting method, denoted as AFQMC/NOCI substantially reduces phaseless errors and the AFQMC sampling variance.
Reducing sampling variance often offsets the additional computational cost introduced by the NOCI trial wave function.

The remainder of the paper is structured as follows:
we briefly recapitulate the AFQMC and NOCI methods and describe the selection process in detail in Sec.~\ref{sec:theory}.
We dedicate Sec.~\ref{sec:compsetup} to the practical implementation details of the selection process that yields the most compact NOCI expansion.
We validate our approach through various applications on small molecular systems in Sec.~\ref{sec:results}.
Finally, Sec.~\ref{sec:conclusion} summarizes our work and possibilities for future developments.

\section{Theoretical Background} 
\label{sec:theory}
We begin this section by describing the AFQMC random walk (Sec.~\ref{subsec:afqmc}).
Then we describe the NOCI method and the selection process in Sec.~\ref{subsec:noci}.

\subsection{AFQMC Random Walk}
\label{subsec:afqmc}
The AFQMC algorithm extracts the many-body ground state $\ket{\Phi}$ from the initial state $\ket{\Phi_0}$ by applying long imaginary time propagation
\begin{equation}
    \label{eq:phiex}
    \ket{\Phi} = \lim_{k \to \infty} \ket{\Phi_k} = \lim_{k \to \infty} \left[ e^{-\tau \left( \hat H - E_0 \right) }  \right]^{k} \ket{\Phi_0},
\end{equation}
where $\tau$ is the time step, $E_0$ is the best estimate of the ground state energy, and $\hat H$ is the many-body Hamiltonian.
At a given time step $k$, we approximate the ground state wave function $\ket{\Phi_k}$ as the weighted average of $N_w$ walkers
\begin{equation}
    \label{eq:psik}
    \ket{\Phi_k} = \Bigl(\sum_{w} W_{k}^{w} e^{i \theta_{k}^{w}}\Bigr)^{-1} \sum_{w} W_{k}^{w} e^{i \theta_{k}^{w}} \frac{\ket{\Psi_{k}^{w}}}{\braket{\Phi_{\guide} | \Psi_{k}^{w}}},
\end{equation}
where $\ket{\Phi_\guide}$ is a trial wave function, and $w$ runs over $N_w$ walkers.
Each walker is represented by a single Slater determinant $\ket{\Psi_k^w}$, a real-valued weight $W_k^w$ and a phase $\theta_k^w$, initialized as
\begin{align}
    W_{0}^w e^{i\theta_0^w} & = \braket{\Phi_\guide | \Psi_\init}; 
    &
    \ket{\Psi_{0}^w} & = \ket{\Psi_{\init}}.
\end{align}
$\ket{\Psi_\init}$ is the initial determinant, usually chosen to be the Hartree-Fock determinant.
With these ingredients, the AFQMC random walk is defined as 
\begin{gather}
    \ket{\Psi_{k+1}^w} = \hat B(\mathbf{x}^w) \ket{\Psi_{k}^w},    \label{eq:psiupdate}    \\
    W_{k+1}^w e^{i \theta_{k+1}^w} = W_{k}^w e^{i \theta_{k}^w}  \; \frac{\braket{\Phi_{\guide} | \Psi_{k+1}^w}}{\braket{\Phi_{\guide} | \Psi_{k}^w}}  I(\mathbf{x}^w).    \label{eq:wupdate}
\end{gather}

The explicit form of the propagator $\hat B(\mathbf{x})$ depends on the specific realization of the Trotter decomposition \cite{Trotter1959} and the Hubbard-Stratonovich transformation.\cite{Stratonovich1957,Hubbard1959}
We showed in our recent work \cite{SukurmaLargeTauAFQMC2024} that
\begin{equation}
\label{eq:afqmc_prop}
    \hat B(\mathbf{x}) = e^{\tau E_0} e^{-\tau \hat H_1/2} e^{i\sqrt{\tau}\sum_g (x_g-f_g) \hat L_g} e^{-\tau \hat H_1 /2}~.
\end{equation}
yields minimal time-step errors.
The vector $\mathbf{x} = \{x_g\}_{g=1}^{N_g}$ is a random vector drawn from the standard normal distribution, and the vector $\mathbf{f} = \{f_g\}_{g=1}^{N_g}$ represents the centers of the shifted Gaussian distribution.
They are chosen as
\begin{equation}
    f_{g} = - i \sqrt{\tau} \ \frac{\braket{\Psi_{\guide} | \hat L_g | \Psi_k^w} }{\braket{\Psi_{\guide} | \Psi_k^w}}
\end{equation}
to minimize the accumulated phase of the walker.
The importance sampling reweighting factor $I(\mathbf{x})$ is the ratio between standard normal and shifted Gaussian distributions
\begin{equation}
    I(\mathbf{x}) = \exp\Bigl[\mathbf{xf} - \frac12 \mathbf{f}^2\Bigr].
\end{equation}
The energy of the system is the weighted average of the local energies
\begin{equation}
    \label{eq:AFQMCtoten}
    E_0 = \frac{\sum_{kw} W_k^w e^{i\theta_k^w}  E_\loc(\Psi_k^w)}{\sum_{kw} W_k^w e^{i\theta_k^w}},
\end{equation}
where the local energy is computed using non-orthogonal Wick's theorem\cite{Balian1969Wick}
\begin{multline}
\label{eq:locen}
    E_\loc(\Psi) = \frac{\braket{\Psi_{\trial} | \hat H | \Psi}}{\braket{\Psi_{\trial} | \Psi}}  \\
                 = \sum_{pq} h_{pq} G_{pq} + \frac{1}{2} \sum_g \sum_{pqrs} L_{g,pq} L_{g,rs} (G_{pq}G_{rs} - G_{ps}G_{rq}).
\end{multline}
The interstate reduced one-body density matrix $G$ is given by
\begin{equation}
    G_{pq} \equiv G_{pq}(\Psi) = \frac{\braket{\Psi_{\trial} | \hat a_{p}^{\vphantom{\dagger}\dagger} \hat a_q| \Psi}}{\braket{\Psi_{\trial} | \Psi}} = \left[ \Psi (\Psi_{\trial}^{\dagger} \Psi)^{-1}  \Psi_{\trial}^{\dagger} \right]_{qp}.
\end{equation}
The previous two equations apply to the single-determinantal $\ket{\Psi_\trial}$, and can be easily generalized for the multi-determinantal wave function $\ket{\Phi_\trial}$.\cite{Shi2021RecentDevAFQMC}

\subsection{NOCI Selection Process}
\label{subsec:noci}
Following from Eqs.~\eqref{eq:phiex}, \eqref{eq:psik}, and \eqref{eq:psiupdate}, the exact ground-state wave function is approximated by an integral over random fields
\begin{equation}
    \ket{\Phi} \approx \int \mathrm{d}^{N_g}x \ c(\mathbf{x}) \hat B(\mathbf{x}) \ket{\Psi_0}, 
\end{equation}
where $c(\mathbf{x})$ is an unknown amplitude function representing $\ket{\Phi}$, and $\hat B(\mathbf{x}) \ket{\Psi_0}$ denotes non-orthogonal Slater determinants for different $\mathbf{x}$ vectors.
Here, $\ket{\Psi_0}$ is a reference determinant, usually a Hartree-Fock wave function.
This approximation would become an equality if $\hat B(\mathbf{x}^1) \hat B(\mathbf{x}^2) = \hat B(\mathbf{x}^1 + \mathbf{x}^2)$.
The discretized form of the previous equation serves as the basis for the NOCI expansion
\begin{equation}
    \ket{\Phi_\noci} = \sum_{\alpha}^{N_{\mathrm{d}}} c_\alpha \hat B(\mathbf{x}^\alpha) \ket{\Psi_0}
    = \sum_{\alpha}^{N_{\mathrm{d}}} c_\alpha \ket{\Psi_\alpha}.
\end{equation}

The selection process extracts $N_{\mathrm{d}}$ Slater determinants $\ket{\Psi_\alpha}$ from the AFQMC random walk and determines the optimal coefficients $c_\alpha$ solving the NOCI equation
\begin{equation}
    \label{eq:nocieq}
    \sum_\beta H_{\alpha\beta} c_\beta = E \sum_\beta S_{\alpha\beta} c_\beta,
\end{equation}
where $E$ is an upper bound to the true ground state energy $E_0$.
The Hamiltonian and the overlap matrix elements
\begin{align}
    H_{\alpha\beta} & = \braket{\Psi_\alpha|\hat H | \Psi_\beta },
    &
    S_{\alpha\beta} & = \braket{\Psi_\alpha|\Psi_\beta},
\end{align}
are computed using non-orthogonal Wick's theorem,\cite{Balian1969Wick} similar to Eq. \eqref{eq:locen}.

We divide the selection process into three stages:
(1) preselection based on AFQMC local energies,
(2) a metric test, and
(3) an energy test, both typically employed in NOCI.\cite{DuttaNOAGP2021,SunSNOCISD2024}

\subsubsection{AFQMC Preselection}
\label{subsubsec:presel}
AFQMC walkers with low negative local energies tend to have large weights, small overlap with the trial wave function $\ket{\Psi_{\trial}}$, and significant contribution to the total AFQMC energy (Eq.~\eqref{eq:AFQMCtoten}).
This makes them suitable candidates for inclusion in the NOCI expansion.
For a given ensemble of walkers in $\ket{\Phi_k}$ defined in Eq.~\eqref{eq:psik}, we calculate the mean local energy
\begin{equation}
    \bar E_{\loc} = \frac{1}{N_w} \sum_w E_{\loc}(\Psi_w)
\end{equation}
and the standard deviation
\begin{equation}
    \sigma_{E_{\loc}}^{2} = \frac{1}{N_w-1} \sum_{w} \bigl( E_{\loc}(\Psi_w) - \bar E_{\loc} \bigr)^2.
\end{equation}
Walkers with local energies satisfying
\begin{equation}
    \label{eq:lambda}
    E_{\loc}(\Psi_w) \leq \bar E_{\loc} - \lambda \sigma_{E_{\loc}}
\end{equation}
are advanced to the further NOCI selection.
$\lambda$ is a numeric parameter that needs to be adjusted to obtain an optimized trial state.
This step requires only local energy information, making it computationally simpler than the subsequent two tests.

\subsubsection{Metric Test}
\label{subsubsec:mtest}
We define the projector operator 
\begin{equation}
    \label{eq:proj}
    \hat Q = 1 - \sum_{\alpha,\beta=1}^{N_d} \ket{\Psi_\alpha} S_{\alpha\beta}^{-1} \bra{\Psi_\beta}
\end{equation}
that projects the candidate determinant $\ket{\Psi}$ on the space orthogonal to the space spanned by $N_{\mathrm{d}}$ determinants in $\ket{\Phi_\noci}$.
We note that the projector operator $\hat Q$ is Hermitian $\hat Q^{\dagger} = \hat Q$ and idempotent $\hat Q^2 = \hat Q$.
The projection $\hat Q \ket{\Psi}$ is not a single Slater determinant and can be interpreted as a residual vector to the space spanned by determinants in $\ket{\Phi_\noci}$.
We define a metric threshold $\mu$ and require 
\begin{equation}
    \label{eq:mtest}
    \sqrt{\frac{\braket{\Psi|\hat Q| \Psi}}{\braket{\Psi|\Psi}}} \geq \mu.
\end{equation}

\subsubsection{Energy Test}
\label{subsubsec:etest}
The energy test measures the contribution of the candidate determinant $\ket{\Psi}$ to the total energy of $\ket{\Phi_\noci}$.
We could compute the exact contribution with the NOCI equation~\eqref{eq:nocieq}, but solving it for every candidate determinant is computationally too expensive.
Instead, we determine coefficients of a smaller variational problem with only two degrees of freedom
\begin{equation}
    \ket{\Phi_{\noci + 1}} = \tilde{c}_1 \ket{\Phi_\noci} + \tilde{c}_2 \hat Q \ket{\Psi}
\end{equation}
by solving 
\begin{equation}
    \label{eq:2nocieq}
    \begin{pmatrix}
    \braket{\Phi_\noci | \hat H | \Phi_\noci} & \braket{\Phi_\noci | \hat H \hat Q | \Psi } \\
    \braket{\Psi | \hat Q \hat H| \Phi_\noci} & \braket{\Psi|\hat Q \hat H \hat Q|\Psi}
    \end{pmatrix}
    \begin{pmatrix}
        \tilde c_1\\
        \tilde c_2
    \end{pmatrix}
    = \tilde E
    \begin{pmatrix}
        \braket{\Phi_\noci | \Phi_\noci} \tilde c_1\\
        \braket{\Psi|\hat Q |\Psi} \tilde c_2
    \end{pmatrix}.
\end{equation}
The variational energy $\tilde{E}$ is the approximation of the full variational energy, and it is sufficiently accurate for the selection process.
A candidate determinant $\ket{\Psi}$ passes the energy test if 
\begin{equation}
    \frac{E-\tilde{E}}{|E|} > \varepsilon
\end{equation}
for an energy threshold $\varepsilon$.

If a candidate determinant $\ket{\Psi}$ meets all three criteria, it is added to the NOCI expansion, and the NOCI equation~\eqref{eq:nocieq} is solved again to update the coefficients $c_\alpha$.

\section{Implementation Details}
\label{sec:compsetup}
In this section, we study the impact of all parameters on the efficiency of the optimized NOCI selection algorithm.
We benchmark the algorithm on the O$_2$ molecule using the cc-pVDZ basis set and frozen-core approximation, a representative system for which AFQMC/HF performs poorly.\cite{sukurmaHEAT2023}

Inspired by Chen \emph{et. al},\cite{ChenHybridAFQMC2023} we constrain the AFQMC random walk during the NOCI selection process to the first 100 Cholesky vectors for all systems considered in this work.
This is achieved by setting $\hat L_g = 0 $ for $g>100$ in Eq.~\eqref{eq:afqmc_prop}.
While not essential, this constraint accelerates the selection process and produces more compact and accurate NOCI wave functions.

We optimize the trial wave function over $\epochs$ epochs.
From one epoch to the next, we systematically decrease the energy threshold $\varepsilon$
\begin{equation}
    \varepsilon_\epoch = \varepsilon_{\mathrm{max}} \Bigl(\frac{\varepsilon_{\mathrm{min}}}{\varepsilon_{\mathrm{max}}}\Bigr)^{(\epoch - 1)/(\epochs - 1)}
\end{equation}
for $\epoch = 1, 2, \ldots, \epochs$.
As a consequence, we add only determinants with very large impact on the energy in early epochs.
In later epochs, where the trial wave function is already more accurate, we also add determinants with a smaller impact on the energy.
In each epoch, we find that sufficiently many non-orthogonal Slater determinants need to be considered.
Beyond that, the accuracy does not improve, so a fixed number of walkers $N_w = 6,400$, number of time steps of 100, and time step $\tau = $\qty{0.05}{\hartree^{-1}} is acceptable.
After each time step, we update the NOCI wave function with determinants that meet all three criteria \ref{subsubsec:presel} - \ref{subsubsec:etest}.
At the end of the epoch, we update the trial wave function for the next one.

\begin{figure}
    \centering
    \includegraphics[width=\columnwidth]{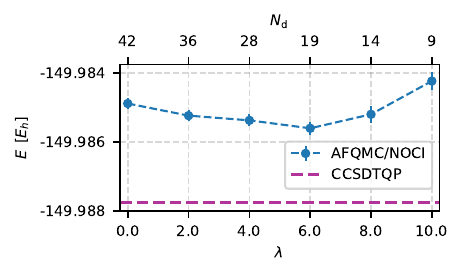}
    \caption{
    AFQMC/NOCI energy (blue dots) does not depend strongly on the preselection parameter $\lambda$, with the deviation from the reference CCSDTQP energy (magenta line) being nearly constant.
    Small $\lambda$ values increase the number of determinants $\numdet$, whereas large values yield a larger sample variance.
    We select $\lambda = 4.0$ as a balanced choice between accuracy and efficiency.
    (Note: The system under consideration is O$_2$ in the cc-pVDZ basis set.
    Other selection parameters are chosen as specified in Table~\ref{tab:params}, with $\varepsilon_{\mathrm{min}}=10^{-5}$.)
    }
    \label{fig:O2lambda}
\end{figure}

\begin{figure}
    \centering
    \includegraphics[width=\columnwidth]{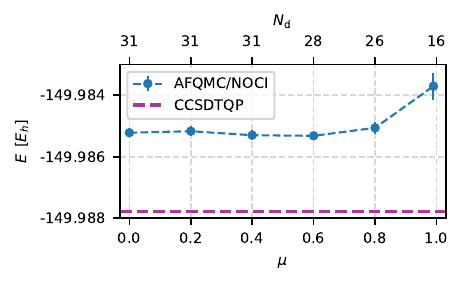}
    \caption{
    AFQMC/NOCI energy (blue dots) converges rapidly as the metric parameter $\mu$ decreases.
    The number of determinants $\numdet$ remains nearly constant as long as $\mu$ is not too large.
    We use $\mu = 0.6$ throughout this work.
    (Note: The system under consideration is O$_2$ in the cc-pVDZ basis set.
    Other selection parameters are chosen as specified in Table~\ref{tab:params}, with $\varepsilon_{\mathrm{min}}=10^{-5}$.)
    }
    \label{fig:O2mu}
\end{figure}

The parameters $\lambda$ for the preselection and $\mu$ for the metric test are constant across all epochs because they have a smaller impact on the accuracy of the method.
Let us consider $\lambda$ first.
Fig.~\ref{fig:O2lambda} shows the AFQMC energy with a NOCI trial wave function (AFQMC/NOCI energy) as a function of the parameter $\lambda$ for otherwise fixed parameters.
Small $\lambda$ values increase the number of determinants without significant improvement of the energy.
Conversely, large $\lambda$ values are overly restrictive and may compromise selection accuracy.
We find $\lambda = 4.0$ to be a balanced choice, selecting about one to ten percent of the walkers as candidates.

Next, we consider the dependence of the AFQMC/NOCI energy on the parameter $\mu$ (Fig.~\ref{fig:O2mu}).
For a wide range of values, $\mu$ has little impact on the number of determinants and the AFQMC/NOCI energy.
Too large values of $\mu$ remove too many candidates leading to a significant increase in the variance.
In other systems, we also observed that small values of $\mu$ may lead to linear-dependency issues.
We set $\mu = 0.6$ as a balanced compromise.

\begin{figure}
    \centering
    \includegraphics[width=\columnwidth]{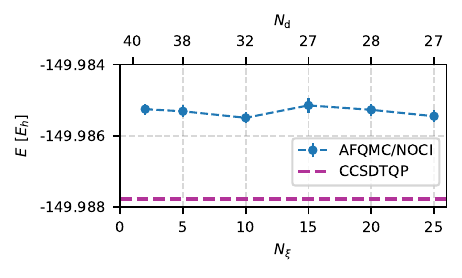}
    \caption{
    AFQMC/NOCI energy (blue dots) depends weakly on the number of epochs $\epochs$ for fixed $\varepsilon_{\mathrm{min}}$, $\varepsilon_{\mathrm{max}}$ values.
    However, the number of determinants $\numdet$ decreases as $\epochs$ increases.
    For typical $\varepsilon_{\mathrm{min}}$ and $\varepsilon_{\mathrm{max}}$ values, we use $\epochs = 10$ throughout this work.
    (Note: The system under consideration is O$_2$ in the cc-pVDZ basis set.
    Other selection parameters are chosen as specified in Table~\ref{tab:params}, with $\varepsilon_{\mathrm{min}}=10^{-5}$.)
    }
    \label{fig:o2epoch}
\end{figure}

Introducing epochs to dynamically adjust the energy test increases the number of adjustable parameters of the selection process.
However, as we will show this effort is justified because the lower bound $\varepsilon_{\mathrm{min}}$ for the energy is the single most important convergence parameter.
The upper bound $\varepsilon_{\mathrm{max}}$ used in the first epoch and the number of epochs $\epochs$ do not significantly alter the results.
$\varepsilon_{\mathrm{max}}$ should be large enough that only few determinants are selected in the first epoch.
For all systems considered in this work, we find that $\num{e-4} \le \varepsilon_{\mathrm{max}} \le \num{4e-4}$
is an appropriate choice.

The number of epochs $\epochs$ controls the final number of determinants in $\ket{\Psi_\noci}$ for fixed $\varepsilon_{\mathrm{min}}$ and $\varepsilon_{\mathrm{max}}$ values. 
Fig.~\ref{fig:o2epoch} shows the AFQMC/NOCI energy as a function of $\epochs$. 
For a small number of epochs, $\varepsilon$ decreases rapidly towards $\varepsilon_\mathrm{min}$ so that more determinants are selected.
As the number of epochs increases, the number of determinants decreases without compromising accuracy. 
Beyond a certain $\epochs$ value, the number of determinants no longer changes, while the computational cost increases linearly with $\epochs$.
As a reasonable compromise, we fix $\epochs = 10$ throughout this work.

\begin{table}
\caption{\label{tab:params}
    List of optimal NOCI selection parameter values used throughout this work, unless stated otherwise.
    The $\varepsilon_{\mathrm{min}}$ parameter is the only remaining adjustable parameter that determines the accuracy of the NOCI selection.
    }
\begin{ruledtabular}
\begin{tabular}{ccccccc}
$N_w$  & $N_k$ & $\tau$                    & $\lambda$ & $\mu$ & $\varepsilon_{\mathrm{max}}$ & $\epochs$  \\ \hline
6\,400   & 100   & 0.05\si{\hartree^{-1}}    & 4.0       & 0.6   & $\geq 10^{-4}$             & 10
\end{tabular}
\end{ruledtabular}
\end{table}

\begin{figure}
    \centering
    \includegraphics[width=\columnwidth]{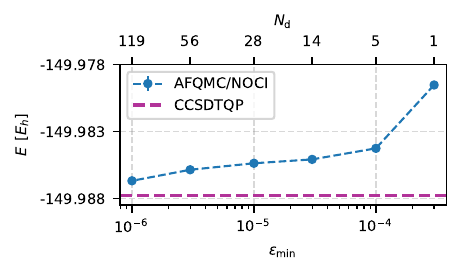}
    \caption{
    AFQMC/NOCI energy (blue dots) converges toward the reference CCSDTQP energy (magenta line) as $\varepsilon_{\mathrm{min}}$ decreases. 
    Simultaneously, the number of determinants $\numdet$ increases with decreasing $\varepsilon_{\mathrm{min}}$.
    (Note: The system under consideration is O$_2$ in the cc-pVDZ basis set.
    Other selection parameters are chosen as specified in Table~\ref{tab:params}.)
    }
    \label{fig:O2emin}
\end{figure}

The remaining tunable parameter $\varepsilon_{\mathrm{min}}$ controls the accuracy of the trial wave function.
For all other parameters, we use the values listed in Table~\ref{tab:params}.
Fig.~\ref{fig:O2emin} illustrates the systematic convergence of the AFQMC/NOCI result to the reference value as $\varepsilon_{\mathrm{min}}$ decreases.
At $\varepsilon_{\mathrm{min}} = 10^{-6}$, the trial wave function contains 119 determinants, reduces the error by \qty{7.1}{\milli \hartree} and agrees with the reference value within chemical accuracy.

\begin{figure}
    \centering
    \includegraphics[width=\columnwidth]{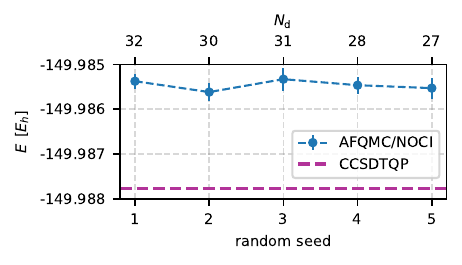}
    \caption{
    AFQMC/NOCI energy does not depend on the randomness of the NOCI selection process.
    Similarly, the number of determinants $\numdet$ reamins nearly constant across all five random seeds used to initialize the AFQMC random walk.
    (Note: The system under consideration is O$_2$ in the cc-pVDZ basis set.
    Other selection parameters are chosen as specified in Table~\ref{tab:params}, with $\varepsilon_{\mathrm{min}}=10^{-5}$.)
    }
    \label{fig:O2seeds}
\end{figure}

Finally, we demonstrate that the randomness of selecting the trial wave function does not impact its accuracy.
We perform five identical selections with $\varepsilon_{\mathrm{min}}=$ \num{e-5}, each using a different random seed.
Fig.~\ref{fig:O2seeds} shows the AFQMC energies and their corresponding standard deviations for each random seed. 
The AFQMC energies are consistent within statistical errors, and the standard deviations remain nearly identical across all random seeds.
Although the final number of determinants in the trial wave function varies slightly, as indicated on the top $x$-axis of Fig.~\ref{fig:O2seeds}, this variation does not affect the accuracy or efficiency of the AFQMC simulation.

\section{Results and Discussion}
\label{sec:results}

In this section, we report AFQMC results using NOCI trial wave functions for the following cases:
(i) second-row atoms, where the Hartree-Fock (HF) trial wave function performs poorly,
(ii) HEAT set molecules,\cite{Tajti2004Heat,Bomble2005Heat,sukurmaHEAT2023}
(iii) the benzene molecule\cite{Benzene2020Blindtest}--- system with a dominant dynamic correlation, and
(iv) the N$_2$ dissociation as an example of strong static correlation effects.

We conduct all HF, NOCI, and AFQMC calculations using the QMCFort code\cite{sukurmaHEAT2023} with the cc-pVDZ basis set and the frozen-core approximation.
For open-shell systems, we use the UHF (unrestricted HF) ground state in both NOCI selection and AFQMC calculations unless explicitly stated otherwise.
The numerical parameters of the NOCI selection were described in detail in the previous section.
We use modified Cholesky decomposition\cite{Beebe1977CholDec,Koch2003CholDec,AquilanteMOLCAS} with a threshold of $10^{-6}$ to represent ERIs.
Our large time-step algorithm\cite{SukurmaLargeTauAFQMC2024} enables a time step of \qty{0.02}{\hartree^{-1}} without residual time-step errors.
All AFQMC calculations employ 6,400 walkers and run until the statistical error is below \qty{0.2}{\milli\hartree}.
Consequently, figures showing AFQMC energies have no visible error bars.
All AFQMC values are provided in the Supporting Inforamtion.

\subsection{2nd Row Atoms}
\label{subsec:atoms}

While CCSD(T) energies for second-row isolated atoms are nearly exact, AFQMC with a HF trial wave function (AFQMC/HF) exhibits unexpectedly large errors.\cite{Borda2019NONSD,sukurmaHEAT2023}
Therefore, single-determinant AFQMC could only predict quantities that do not involve the atomic energies of these elements.
Consequently, these systems provide an ideal starting point for assessing the quality of the AFQMC using NOCI trial wave functions.
We exclude the Li atom for which HF theory is already exact in the frozen-core approximation.

\begin{figure}
    \centering
    \includegraphics[width=\columnwidth]{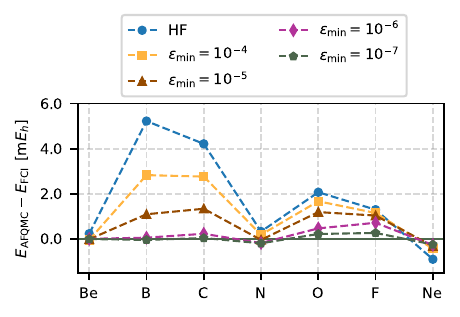}
    \caption{
    For second-row atoms, AFQMC/NOCI energies converge from AFQMC/HF toward the exact FCI energies as one tightens $\varepsilon_{\mathrm{min}}$.
    (Note: Calculations are performed using the cc-pVDZ basis set and frozen-core approximation.)
    }
    \label{fig:atoms}
\end{figure}

\begin{table}
\caption{\label{tab:atoms}
    Statistics for second-row elements for AFQMC/HF and AFQMC/NOCI at various $\varepsilon_{\mathrm{min}}$ values, with the cc-pVDZ basis set and frozen-core approximation.
    Root-mean-square deviation (RMSD), mean absolute deviation (MAD), and maximal absolute deviation $\max(|\Delta E|)$ are presented in \si{\hartree} units.
    $\langle\numdet\rangle$ is the average number of determinants for each  $\varepsilon_{\mathrm{min}}$ value and  $\max(\numdet)$ is the maximum number of determinants across all atoms for a given $\varepsilon_{\mathrm{min}}$ value.
}
\begin{ruledtabular}
\begin{tabular}{lccccc}
$\varepsilon_{\mathrm{min}}$ & RMSD   & MAD  & $\max(|\Delta E|)$  & $\langle\numdet\rangle$       & $\max(\numdet)$  \\ \hline
HF                           & 2.7   & 2.0    & 5.2                 & 1             & 1   \\
$10^{-4}$                    & 1.7   & 1.3    & 2.8                 & 5             & 9   \\
$10^{-5}$                    & 0.9   & 0.7    & 1.3                 & 25            & 34  \\
$10^{-6}$                    & 0.4   & 0.3    & 0.7                 & 82            & 111  \\
$10^{-7}$                    & 0.2   & 0.1    & 0.3                 & 129           & 179
\end{tabular}
\end{ruledtabular}
\end{table}

Fig.~\ref{fig:atoms} depicts atomic AFQMC energies relative to FCI energies for the HF trial determinant and NOCI trial wave functions at several $\varepsilon_{\mathrm{min}}$ values. 
The AFQMC energies systematically converge toward the FCI values as $\varepsilon_{\mathrm{min}}$ decreases.
The root-mean-square deviation (RMSD) reduces from \qty{2.7}{\milli \hartree} for the HF trial wave function to within chemical accuracy for the NOCI trial wave function at $\varepsilon_{\mathrm{min}} = 10^{-5}$.
At this threshold, the NOCI selection yields an average of 25 Slater determinants.
Tighter thresholds of $\varepsilon_{\mathrm{min}} = 10^{-6}$ or $\varepsilon_{\mathrm{min}} = 10^{-7}$ reduce the error to \qty{0.4}{\milli \hartree} or \qty{0.2}{\milli \hartree} with
an average of 82 or 129 Slater determinants, respectively.
More detailed statistics are provided in Table~\ref{tab:atoms}.

\subsection{HEAT Set}
\label{subsec:HEAT}
The HEAT dataset\cite{Tajti2004Heat,Bomble2005Heat} serves as a reliable benchmark for testing state-of-the-art correlation-consistent methods.
We compare nearly exact CCSDTQP energies to AFQMC/HF, AFQMC/NOCI, and CCSD(T) ones. 
Although the largest system in the HEAT set (CO$_2$ molecule) consists of 16 electrons in 39 orbitals, both CCSD(T) and AFQMC/HF have a relatively high RMSD of \qty{1.7}{\milli \hartree} and \qty{2.9}{\milli \hartree}, respectively.
This could be attributed to the fact that 14 systems are open-shell molecules, which are generally more challenging for quantum chemistry methods.

\begin{figure*}
    \centering
    \includegraphics[width=0.9\textwidth]{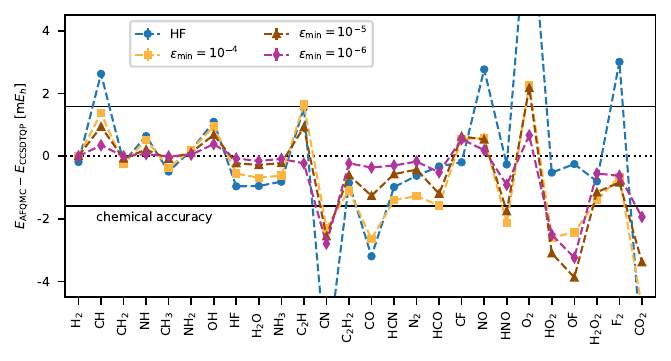}
    \caption{
    For HEAT set molecules, AFQMC/NOCI energies converge toward the CCSDTQP values as we tighten $\varepsilon_{\mathrm{min}}$.
    The RMSD decreases from \qty{2.8}{\milli \hartree} for AFQMC/HF to \qty{1.1}{\milli \hartree} for AFQMC/NOCI at $\varepsilon_{\mathrm{min}} = 10^{-6}$.
    (Note: Calculations are performed using the cc-pVDZ basis set and frozen-core approximation.)
    }
    \label{fig:heat}
\end{figure*}

We calculate AFQMC/HF and AFQMC/NOCI energies at three different $\varepsilon_{\mathrm{min}}$ values. 
Fig.~\ref{fig:heat} illustrates the errors in AFQMC/HF and AFQMC/NOCI energies at different $\varepsilon_{\mathrm{min}}$ values relative to the CCSDTQP values.
The RMSD decreases from \qty{2.8}{\milli \hartree} for AFQMC/HF to \qty{1.5}{\milli \hartree} at $\varepsilon_{\mathrm{min}} = 10^{-5}$, and further to \qty{1.1}{\milli \hartree} at $\varepsilon_{\mathrm{min}} = 10^{-6}$.
The average number of determinants in these calculations is 35 and 194, respectively.
It is noteworthy that the residual errors are mainly associated with open-shell molecules. 
To demonstrate this, we calculate the RMSD for a subset of closed-shell HEAT molecules.
For AFQMC/NOCI at $\varepsilon_{\mathrm{min}} = 10^{-6}$, the RMSD is reduced to \qty{0.7}{\milli \hartree} compared to \qty{2.3}{\milli \hartree} for AFQMC/HF.
More detailed statistics are given in table~\ref{tab:heat}.

The CN molecule is a special case in the HEAT set.
Since CN exhibits a large spin contamination, starting from an unrestricted HF determinant leads to larger AFQMC errors than using a restricted one.
Hence, we start the NOCI selection for CN from the RHF (restricted HF) determinant.
We will study this behavior further in the section on $N_2$ dissociation (Sec.~\ref{subsec:n2dis}).

\begin{table}
\caption{\label{tab:heat}
    Statistics for HEAT set molecules for AFQMC/HF and AFQMC/NOCI at various $\varepsilon_{\mathrm{min}}$ values, with the cc-pVDZ basis set and frozen-core approximation.
    Root-mean-square deviation (RMSD), mean absolute deviation (MAD), and maximal absolute deviation $\max(|\Delta E|)$ are presented in \si{\hartree} units.
    $\langle\numdet\rangle$ is the average number of determinants for each  $\varepsilon_{\mathrm{min}}$ value and  $\max(\numdet)$ represents the maximum number of determinants across all HEAT molecules for a given $\varepsilon_{\mathrm{min}}$ value.
}
\begin{ruledtabular}
\begin{tabular}{lccccc}
$\varepsilon_{\mathrm{min}}$ & RMSD   & MAD  & $\max(|\Delta E|)$  & $\langle\numdet\rangle$       & $\max(\numdet)$  \\ \hline
HF                           & 2.8    & 1.7  &  8.5                & 1             & 1   \\
$10^{-4}$                    & 1.7    & 1.3  &  4.8                & 6             & 30   \\
$10^{-5}$                    & 1.5    & 1.1  &  3.9                & 35            & 59  \\
$10^{-6}$                    & 1.1    & 0.7  &  3.2                & 194           & 324 
\end{tabular}
\end{ruledtabular}
\end{table}

\subsection{N$_2$ Dissociation}
\label{subsec:n2dis}
The dissociation of the N$_2$ molecule is a well-known example of strong static correlation effects caused by the breaking of the triple bond.
The system also undergoes a symmetry-breaking transition, with a molecular-like RHF ground state at equilibrium geometry and an atomic-like UHF ground state at stretched geometries.
The latter exhibits strong spin contamination and poses a significant challenge for AFQMC/NOCI.

We compute AFQMC total energies at six different bond lengths $R$, ranging from the equilibrium bond length of \qty{2.118}{\bohr} to \qty{4.2}{\bohr}.
We report AFQMC results for three trial wave functions:
(i) the UHF determinant, denoted as AFQMC/UHF;
(ii) the NOCI wave function selected through the AFQMC random walk, denoted as AFQMC/NOCI; and
(iii) a special NOCI wave function obtained by diagonalizing only the RHF and two UHF determinants, denoted as AFQMC/RHF-UHF.
Two UHF determinants are constructed by swapping the spin-up and spin-down orbitals.
This trial wave function is motivated by the observation that both molecular and atomic-like solutions contribute significantly to the ground state near bond dissociation.

\begin{figure}
    \centering
    \includegraphics[width=\columnwidth]{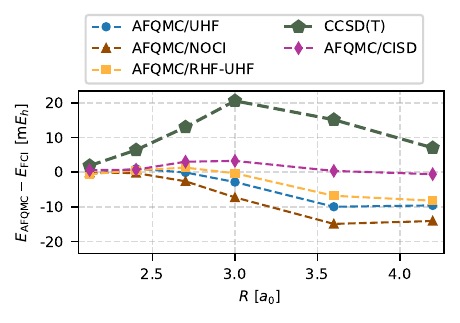}
    \caption{
    CCSD(T) energies and AFQMC energies with different trial wave functions, shown relative to the FCI energies.
    AFQMC/NOCI performs worse than AFQMC/UHF for UHF trial wave functions with strong spin contamination.
    AFQMC/RHF-UHF yields nearly exact energies at intermediate bond lengths, where CCSD(T) and AFQMC/NOCI exhibit their largest errors.
    (Note: Exact FCI energies are taken from Ref.~\onlinecite{ChanDMRGN2dis2004}, and AFQMC/CISD energies are taken from Ref.~\onlinecite{MahajanCISD-AFQMC2024}.
    All calculations are performed using the cc-pVDZ basis set and frozen-core approximation.)
    }
    \label{fig:n2dis}
\end{figure}

Fig.~\ref{fig:n2dis} shows the errors in our AFQMC energies, CCSD(T) energies, and recently published AFQMC/CISD energies.\cite{MahajanCISD-AFQMC2024}
The reference values are exact DMRG energies taken from Ref.~\onlinecite{ChanDMRGN2dis2004}.
CCSD(T) performs poorly with a maximal error of \qty{20.6}{\milli \hartree}.
Our AFQMC/UHF energies agree well with those published in Ref.~\onlinecite{MahajanCISD-AFQMC2024}, yielding a maximal error of \qty{10.0}{\milli \hartree}.
Unfortunately, our AFQMC/NOCI results are even less reliable than AFQMC/UHF, with an error of \qty{14.6}{\milli \hartree}.
This confirms the suspicion that the NOCI selection using spin-contaminated UHF determinants leads to unreliable trial wave functions.
We aim to address this issue in future work.
AFQMC/RHF-UHF achieves nearly exact results at $ R= $ \qty{2.7}{\bohr} and $R = $ \qty{3.0}{\bohr}, where both CCSD(T) and AFQMC/CISD exhibit their largest errors.
This is an advantage of the AFQMC method, where compact trial wave functions, straightforwardly constructed using chemical intuition, outperform more sophisticated approaches that rely on the brute-force inclusion of determinants.
However, AFQMC/RHF-UHF still fails at larger bond lengths, where the CI coefficient of the RHF determinant practically drops to zero.
As a result of these large errors at stretched geometries, AFQMC/RHF-UHF produces errors up to \qty{8.2}{\milli \hartree}.

\subsection{Benzene Molecule}
\label{subsec:benzene}
The ground state of benzene, a singlet closed-shell configuration without near-degeneracies, is accurately represented by the Hartree-Fock determinant.
However, the delocalized $\pi$-electrons across the six carbon atoms result in highly correlated electron motion in the molecular orbitals, leading to significant dynamic correlation effects that the Hartree-Fock theory completely neglects.
Although the total correlation energy of \qty{-863}{\milli \hartree},\cite{Benzene2020Blindtest} is dominated by dynamic correlation, both CCSD(T) and AFQMC/RHF exhibit notable errors.
Specifically, CCSD(T) undercorrelates by \qty{3.6}{\milli \hartree}, while AFQMC/RHF overcorrelates by \qty{3.2}{\milli \hartree}.\cite{sukurmaHEAT2023} 
This makes benzene an excellent system to demonstrate the capabilities of correlation-consistent quantum chemistry methods.

\begin{figure}
    \centering
    \includegraphics[width=\columnwidth]{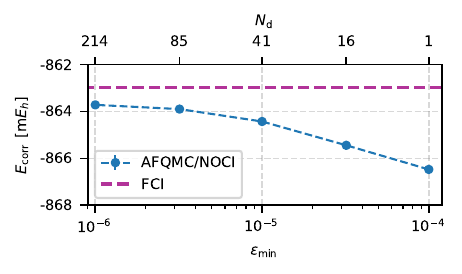}
    \caption{
    The correlation energy of benzene converges smoothly to the FCI value as $\varepsilon_{\mathrm{min}}$ decreases.
    Similarly, the number of determinants $\numdet$ increases with decreasing $\varepsilon_{\mathrm{min}}$, achieving chemical accuracy with only 41 Slater determinants.
    (Note: calculations are performed using the cc-pVDZ basis set and frozen-core approximation.)
    }
    \label{fig:benzene}
\end{figure}

We compute the benzene molecule with 30 electrons in 108 orbitals and compare AFQMC/NOCI to AFQMC/HF.
Fig.~\ref{fig:benzene} illustrates that the AFQMC/NOCI correlation energy converges smoothly towards the FCI limit with tighter thresholds $\varepsilon_{\mathrm{min}}$.
At $\varepsilon_{\mathrm{min}} = 10^{-5}$ (41 Slater determinants), the method reaches chemical accuracy with an error of \qty{1.4}{\milli \hartree}.
This error is reduced further to \qty{0.7}{\milli \hartree} at $\varepsilon_{\mathrm{min}} = 10^{-6}$, where the trial wave function contains 214 determinants.

Interestingly, the number of determinants $\numdet$ required to converge the AFQMC/NOCI energy for the benzene molecule increases roughly linearly with the system size compared to the second-row atoms and HEAT set molecules.
For example, the best NOCI trial wave functions have $\numdet = 65$ for the Ne atom, $\numdet = 135$ for the CO$_2$ molecule, and $\numdet = 214$ for the benzene molecule, which corresponds roughly to 8 determinants per electron for all three systems.
In contrast, the number of determinants in the CISD trial wave function increase from about $\numdet=2000$ for the Ne atom, to $\numdet = 88000$ for the CO$_2$ molecule, and $\numdet = 2.8$~million for the benzene molecule.
A more detailed analysis of how the number of determinants in the NOCI trial wave function scales with system size will be the subject of future work.

Furthermore, we analyze how the AFQMC/NOCI walltime and AFQMC/NOCI error vary with an increasing number of determinants.
We perform this analysis only for the benzene molecule, as it is the largest system considered in this work. 
For second-row atoms and HEAT molecules, such an analysis could lead to misleading conclusions, as the trial wave function approaches the zero variance property, which is not realistic for larger systems of greater interest in real-world applications.

Fig.~\ref{fig:benzene_timing} illustrates how the AFQMC/NOCI walltime depends on the number of determinants in the NOCI trial wave function.
For clarity, the walltime is presented relative to the AFQMC/HF walltime.
The blue line represents the increase in computational cost for the fixed number of AFQMC steps, while the magenta line represents it for the fixed statistical errors.
As the number of determinants increases, the blue line enters the regime where the AFQMC/NOCI walltime scales linearly with the number of determinants.
We expect this behavior becuase the cost of local energy and force bias evaluations scales linearly with the number of determinants. 
However, for fixed statistical errors (magenta line), the AFQMC/NOCI walltime scales as $N_{\mathrm{d}}^{1/4}$ with the number of determinants, demonstrating a more favorable scaling with respect to $N_{\mathrm{d}}$.
As a result, the AFQMC/NOCI with 214 determinants is only eight times slower than AFQMC/HF.
Additional tests are needed to determine whether this behavior generalizes to other systems.

Finally, we examine how the AFQMC/NOCI error varies with the number of determinants, as illustrated in Fig.~\ref{fig:benzene_errors}.
The log-log plot of the AFQMC error versus the number of determinants reveals that the error decreases as $N_{\mathrm{d}}^{-0.6}$, which is similar to the behavior of statistical errors in Monte Carlo simulations.
As mentioned earlier, additional tests are required to confirm these results in general.

\begin{figure}
    \centering
    \includegraphics[width=\columnwidth]{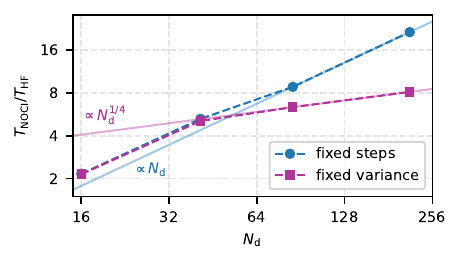}
    \caption{
    The AFQMC/NOCI walltime for a fixed number of AFQMC steps scales linearly with the number of determinants.
    In contrast, for a fixed variance, the scaling becomes sublinear, following a power law with an exponent of 0.25.
    As a result, the AFQMC/NOCI with 214 determinants is only eight times slower than AFQMC/HF.
    The analysis is performed on the benzene molecule using the cc-pVDZ basis set and the frozen-core approximation.
    }
    \label{fig:benzene_timing}
\end{figure}

\begin{figure}
    \centering
    \includegraphics[width=\columnwidth]{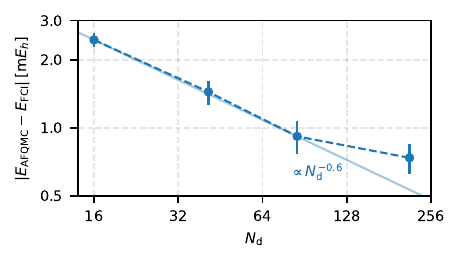}
    \caption{
    The AFQMC/NOCI error decreases approximately as $N_{\mathrm{d}}^{-1/2}$, similar to the stochastic errors in Monte Carlo simulations.
    The analysis is performed on the benzene molecule using the cc-pVDZ basis set and the frozen-core approximation.
    }
    \label{fig:benzene_errors}
\end{figure}

\section{Conclusion}
\label{sec:conclusion}
In this work, we combined the non-orthogonal configuration interaction (NOCI) method with the phaseless auxiliary-field quantum Monte Carlo (AFQMC) random walk to generate more accurate trial wave functions for AFQMC.
We introduced an efficient selection algorithm to identify the most important AFQMC random walkers based on three criteria:
(i) low local energies,
(ii) small overlap with the currently sampled trial wave function, and
(iii) a significant lowering of the variational energy of the trial wave function.
We used the O$_2$ molecule, a system where AFQMC/HF performs poorly, to calibrate the algorithm.
Most parameters of the selection algorithm have minimal impact on its accuracy and can be set to reasonable default values.
In the end, a single adjustable parameter $\varepsilon_{\mathrm{min}}$ governs the accuracy and the number of determinants in the sampled trial wave function.
We showed that AFQMC/NOCI systematically converges to the reference value as we tighten $\varepsilon_{\mathrm{min}}$.

We demonstrated that the NOCI trial wave function improves AFQMC to achieve chemical accuracy for atoms and the HEAT set.
For second-row elements, AFQMC/NOCI reduces the RMSD compared to AFQMC/HF by a factor of 10 using an average of 129 Slater determinants.
Similarly, an average of 35 determinants obtained with $\varepsilon_{\mathrm{min}} = 10^{-5}$ are sufficient to reduce} the RMSD for the HEAT molecules to less than \qty{1}{\kilo \cal / \mol}.
For the benzene molecule, we reduced the AFQMC errors by 80\% with a NOCI trial wave function of 214 determinants. 
In addition, the AFQMC sampling variance is significantly reduced with NOCI wave functions, compensating for the increased computational cost compared to AFQMC/HF.
For the largest system considered in this work, the benzene molecule, 214 determinants increase the computational cost tenfold compared to the corresponding AFQMC/HF calculation. 

Future work should focus on improving the AFQMC/NOCI for strongly correlated systems.
Using N$_2$ dissociation as an example, we have demonstrated that AFQMC/NOCI can sometimes result in larger errors than AFQMC/HF.
A potential solution could be to include orbital rotations in the NOCI optimization using a resonating Hartree-Fock or similar approach.

However, for weakly correlated systems, our selection algorithm, based on AFQMC random walks, provides a systematic approach to refining AFQMC trial wave functions without the need for another approach to determine multi-determinantal trial wave functions.
We demonstrated that trial states of 100-200 non-orthogonal Slater determinants achieve AFQMC energies within chemical accuracy for all weakly correlated systems analyzed in this work. This is particularly exciting for solid-state systems, where sophisticated many-body approaches are often not yet available or computationally still very expensive.

\section{Acknowledgments}
Funding by the Austrian Science Foundation (FWF) within the project P 33440 is gratefully acknowledged. All calculations were performed on the VSC4 / VSC5 (Vienna scientific cluster). 

\section*{Author declarations}
\subsection*{Conflict of Interest}
The authors have no conflicts to disclose.

\section*{Data availability}
The data that support the findings of this study are available within the article.


\bibliographystyle{aapmrev4-1}
\bibliography{bibliography}         

\end{document}